# Rule Based Expert System for Diagnosis of Neuromuscular Disorders

**\*Rajdeep Borgohain**
Department of Computer Science and Engineering, Dibrugarh University Institute of Engineering and Technology, Dibrugarh, Assam
Email: Rajdeepgohain@gmail.com
**Sugata Sanyal**
School of Technology and Computer Science, Tata Institute of Fundamental Research, Mumbai, India
Email: sanyals@gmail.com

**\*Corresponding Author**

-------------------------------------------------------------ABSTRACT--------------------------------------------------------------
In this paper, we discuss the implementation of a rule based expert system for diagnosing neuromuscular diseases. The proposed system is implemented as a rule based expert system in JESS for the diagnosis of Cerebral Palsy, Multiple Sclerosis, Muscular Dystrophy and Parkinson's disease. In the system, the user is presented with a list of questionnaires about the symptoms of the patients based on which the disease of the patient is diagnosed and possible treatment is suggested. The system can aid and support the patients suffering from neuromuscular diseases to get an idea of their disease and possible treatment for the disease.

## 1. INTRODUCTION

With the advances in technology, the domain of computer science is not only limited to its core areas like Computer Networks [1], Network Security [2], Database [3] etc. but it has also transcended to domains like chemistry [4], biology [5], medical diagnosis [6] etc. Computer scientists are finding newer and better ways to apply the knowledge of computer science to other areas. One of the most significant applications of computer science is in the field of medical diagnosis. The knowledge of computer science has been aptly applied to make diagnosis process easier and faster. One such application is the use of expert system to diagnose different diseases.

Expert System is software which can replicate the thinking process of a human and make logical decisions accordingly. Expert systems are used in many areas for automated reasoning. Expert systems finds its place in educational software, decision support systems etc. In this paper, we look at the design and implementation of an expert system for diagnosis of neuromuscular disorders. Expert systems can play a big role in diagnosis of patients with different diseases. The application of expert system to diagnose diseases started in the 70's with the development of Mycin [7] which was developed to detect disease causing bacteria. Since then, many more expert system for medical diagnosis has been developed. Prominent among them are expert systems like INTERNIST – I and CADUCEUS etc. [8]

In this paper, we consider the four of the most common neuromuscular disorder i.e. Cerebral Palsy, Parkinson's disease, Multiple Sclerosis and Muscular Dystrophy and diagnose them with the help of the expert system. The *Expert System for Diagnosis of Neuromuscular Disorders* has a graphical user interface where the user of the system is asked to answer a few questions in yes or no which are prepared according to the symptoms shown by neuromuscular disorder patients. According to the feedback of the user, the expert system uses the RETE algorithm to search the knowledge base and matches the patterns of the symptoms to those in the knowledge base. If the pattern matches, the expert system shows the user the diagnosis of his disease. The system also advises the user to undergo the tests for confirmation of the disease. It also offers the patient with the treatment options.

The rest of the paper is organized in the following way. Section 2 gives an overview of the neuromuscular diseases that the expert system can diagnose. In section 3 we discuss the architecture of the expert system. Section 4 discusses the methodology for diagnosis. In section 5 we look at the implementation of the expert system. In section 6, we discuss the evaluation of the expert system. Finally in section 7, we give the conclusion.

## 2. OVERVIEW OF NEUROMUSCULAR DISEASES

Neuromuscular diseases are those diseases which affects the peripheral nervous system that includes the muscles, the nerve muscle junctions and motor-nerve cells in the spinal cord [9]. Although, neuromuscular diseases encompass lots of diseases, we particularly look at four of the most common neuromuscular diseases, i.e. Cerebral Palsy, Parkinson's disease, Multiple Sclerosis and Muscular Dystrophy.

### 2.1 CEREBRAL PALSY
Cerebral Palsy is a non-progressive disease which affects the posture, gait, movement and muscle tone of the patient. Cerebral Palsy is caused mainly due to injury during fetal stage or at early childhood. Cerebral Palsy is a neuromuscular disease which severely restricts the day to day activities of the patient [10]. According to statistics presented in [11], the number of children suffering from Cerebral Palsy is 1.5 to 4 per 1000 births. The most common symptoms of Cerebral Palsy are:
1. The symptoms appear before 18 months of age.
2. The disease is non-progressive.
3. The disease affects the gait of the patient.
4. The patient is affected by spasticity.

## 2.2 PARKINSON'S DISEASE

Parkinson's disease is a progressive neuromuscular disease which affects the posture, movement, speaking and writing abilities of the patient. Parkinson's disease is caused by gradual deterioration of the nerve cells in the brain with age [12]. Parkinson's disease starts with slight tremors in the patients which hinders with the speech, writing skills and various day to day activities of the patients. Parkinson's disease also affects the posture of the patient and generally the patient walks with a stooping posture. According to [13], more than 60,000 patients are diagnosed with Parkinson's disease in USA alone. The major symptoms of Parkinson's disease are:
1. Postural Instability in the patient.
2. Difficulty in movement of the patient.
3. Difficulty in gait of the patient.
4. Tremors and Shivers in the patient.

## 2.3 MUSCULAR DYSTROPHY

Muscular Dystrophy is a neuromuscular disorder where several muscles groups are affected where the muscles become weak and there is loss of muscle tissue over time. Muscular dystrophy is mainly an inherited disorder, where the disease is passed down the family [14]. The patient of muscular dystrophy suffers from muscle weakness which gets worse over time. The primary symptoms of Muscular dystrophy are:
1. Muscle wasting of patient.
2. Muscle weakness in patient.
3. Difficulty in gait.
4. Difficulty in maintaining balance.

## 2.4 MULTIPLE SCLEROSIS

Multiple Sclerosis is an autoimmune disease that affects the central nervous system. The primary cause of Multiple Sclerosis is the damage of myelin sheath for which the nerve signals slows down [15]. The most common symptoms for multiple sclerosis are:
1. Change in sensation of the patient such as pricking or tingling sensation.
2. Change in strength of the patient.
3. The patient suffers from balance problems.
4. The vision of the patient is affected.

## 3. ARCHITECTURE OF THE PROPOSED SYSTEM

Our proposed system, *Expert System for Diagnosis of Neuromuscular Diseases,* is a rule based expert system which is implemented using Java Expert System Shell (JESS). The system makes use of backward chaining for the inference engine and uses the RETE algorithm to search the knowledge base. The system has a graphical user interface where the user finds two windows, where one is an interactive window and other is the recommendation window. The user is presented with a series of questionnaire in the interactive window which the user has the option to answer in yes or no. The set of questions are prepared according to the symptoms shown by patients of neuromuscular diseases. Now, according to the feedback given by the patient, the RETE algorithm searches the knowledge base for possible pattern matches. If there is a rule in the knowledge base which matches the symptoms of the patient, the system shows the possible diagnosis in the recommendation window. It also advises the patient about the tests to confirm the disease. Moreover, the patient is also given the different treatment options treating the disease.

The architecture of the *Expert System for Diagnosis of Neuromuscular Disorders* is:

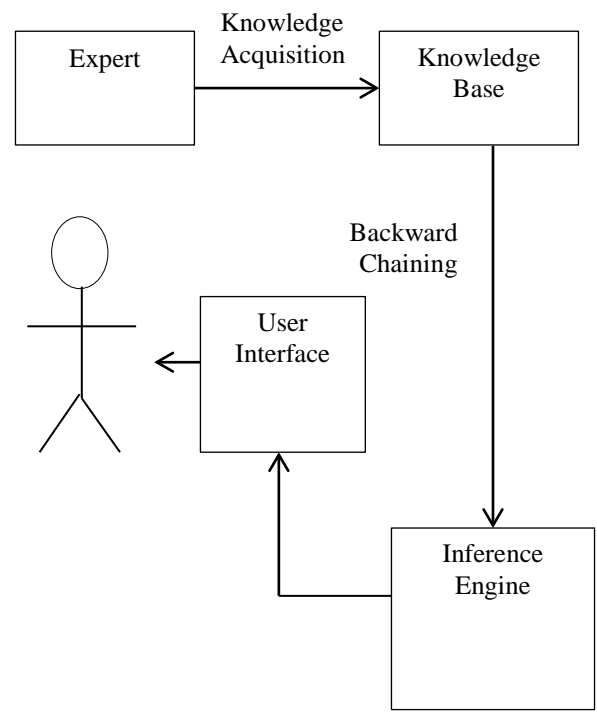

Figure 1. Architecture of the Expert System for Diagnosis of Neuromuscular disorders .

## 3.1 KNOWLEDGE ACQUISITION

The knowledge base can be considered as the heart of the Expert system as all the required facts for building the rules are contained in the knowledge base. Taking this knowledge as the source, rules for the Expert System can be formed [16]. The primary source for knowledge acquisition for the *Expert System for Diagnosis of Neuromuscular Disorders* was consultation with neurology doctors, internet and medical books. The knowledge based consisted of acquiring the symptoms of the disease, treatment and test options etc.

### 3.2 KNOWLEDGE REPRESENTATION

The knowledge base was represented in the form of rules using the Java Expert System Shell. Here the user of the system is first presented with a list of questions which the user has to answer in yes or no. With the help of the questions the symptoms of the disease of the patients are acquired. Then by using the RETE algorithm, which is a pattern matching algorithm, the knowledge base is searched for matching the symptoms of the patients with those already present in the knowledge base. For this, our proposed system uses backward chaining method. Once the system finds the appropriate match, the diagnosis is shown to the user along with the tests that the patient has to perform and the options for treatment.

Let us suppose the patient is suffering from cerebral palsy, the rule for Cerebral Palsy will be:

```
(defrule Cerebral-Palsy
   (declare (auto-focus TRUE))
   (answer (ident progress) (text no))
   (answer (ident age) (text yes))
   (answer (ident gait) (text yes))
   (answer (ident spasticity) (text yes))

   =>
   (recommend-action " The patient is suffering from Cerebral Palsy")
```

Again, if the patient is suffering from Parkinson's disease, the rule will be:

```
(defrule Parkinson
   (declare (auto-focus TRUE))
   (answer (ident posture) (text yes))
   (answer (ident movement) (text yes))
   (answer (ident seizures) (text yes))
   (answer (ident gait) (text yes))

  =>
   (recommend-action "The patient is suffering from Parkinson's disease")
```

For Muscular Dystrophy, we have the following rule:

```
(defrule muscular-dystrophy
   (declare (auto-focus TRUE))
   (answer (ident muscle-wasting) (text yes))
   (answer (ident spasticity) (text yes))
   (answer (ident gait) (text yes))
   (answer (ident balance) (text yes))

 =>
   (recommended-action "The patient is suffering from Muscular Dystrophy")
```

Again, the rule for Multiple Sclerosis is:

```
(defrule multiple-sclerosis
   (declare (auto-focus TRUE))
   (answer (ident sensation) (text yes))
   (answer (ident balance) (text yes))
   (answer (ident vision) (text yes))
   (answer (ident strength) (text yes))

 =>
    (recommend-action "The patient is suffering from Multiple Sclerosis.)
```

### 3.3 THE RETE ALGORITHM

The RETE algorithm is used for searching the knowledge base of the expert system. It is a pattern matching algorithm which drastically reduces the search time by limiting the efforts to re-compute conflicts after a rule is fired [17]. The RETE algorithm is implemented as a directed acyclic graph which is used to match rules to the facts [18].

### 4. METHODOLOGY FOR DIAGNOSIS

.The *Expert System for Diagnosis of Neuromuscular Diseases* implements a decision tree for diagnosis of the diseases. The decision tree acts as a map for the reasoning process which has nodes according to the symptoms. For each question, answered in yes or no, a different node is selected. The system makes use of backward chaining method to reach at the conclusion. Suppose, if a patient is has the symptoms that the disease is non-progressive, the symptoms appeared before 18 months of age, the patient has problem in gait and is suffering from spasticity. Now, the expert system starts with the assumption that the patient is suffering from Cerebral Palsy. On being queried about the progressiveness of the disease, if the patient answers in no, the expert system will discard the assumption that the patient is suffering from Cerebral Palsy. On being answered in yes, the expert system further queries about the age of the patient. If the patient answers that the symptoms appeared after 18 months of age, the system will discard the assumption that the disease in Cerebral Palsy and instead focus on other diseases. On the other hand, if the patient answers in yes, the system will further inquire about the gait and spasticity of the patient. After all the inputs from the patient, the expert system will use the RETE algorithm to search the knowledge base for matching the pattern of the symptoms of the disease. If the pattern matches with Cerebral Palsy, the rule of Cerebral Palsy will be fired and the system will display the diagnosis along with the tests to be done and treatment options.

## 5. IMPLEMENTATION

The proposed *Expert System for Diagnosis of Neuromuscular Disorders* has its inference engine implemented in JESS. There is option for updating the expert system with new knowledge and new rules without changing the whole configuration and adding only the specific rule. As soon as the rule is coded as shown in Section 3.2, the rule is ready and if the pattern matches, the rule will be fired. For the graphical user interface of the system, we have used Java Swing. The graphical user interface provides the user with two windows. One is the interactive window where the questions will be displayed and the user has to answer to the questions. Another is the recommendation window. In the recommendation window, the diagnosis of the patient is shown along with the tests to be done and the treatment options.

## 6. EVALUATION OF THE SYSTEM

The system was tested against some proven cases of neuromuscular disorders. The system showed accurate results when the symptoms of the user were given as input. Moreover, user input from some patients suffering from the diseases were taken and provided as input. The system could diagnose those patients accurately.

## 7. CONCLUSION

In this paper we presented an *Expert System for Diagnosis of Neuromuscular Disorders,* which is used to diagnose some of the most common neuromuscular diseases i.e. Cerebral Palsy, Muscular Dystrophy, Parkinson's disease and Multiple Sclerosis. The system is a rule based expert system implemented using the Java Expert System Shell using the backward chaining mechanism. The expert system can go a great deal in supporting the decision making process of medical professionals and also help patients with Neuromuscular Disorders and give an overview of the disease and treatment options.

### REFERENCES

[1] Bhavyesh Divecha, Ajith Abraham, Crina Grosan and Sugata Sanyal, "Impact of Node Mobility on MANET Routing Protocols Models*", Journal of Digital Information Management*, Volume 5, Number 1, pp. 19-24, 2007

[2] Dhaval Gada, Rajat Gogri, Punit Rathod, Zalak Dedhia, Nirali Mody, Sugata Sanyal and Ajith Abraham, "A Distributed Security Scheme for Ad Hoc Networks", *ACM Crossroads, Special Issue on Computer Security*. Volume 11, No. 1, September, 2004, pp. 1-17

[3] Abraham Silberschatz, Henry F. Korth and S. Sudarshan, *Database System Concepts*, 6th Ed, McGraw Hill, 2010, pp. 17-1340

[4] Adam Roberts, Harold Pimentel, Cole Trapnell and Lior Pachter, "Identification of novel transcripts in annotated genomes using RNA-Sequence", *Bioinformatics*, 27 (2011), pp. 2325--2329.

[5] Stefan Grimme, "Semi empirical GGA–type density functional constructed with a long–range dispersion correction", *Journal of Computational Chemistry*, 2006, 27(15), pp. 1787–1799

[6] Peter R. Innocent and Robert I. John, "Computer aided fuzzy medical diagnosis", *Journal Information Sciences, Special issue: Medical expert systems*, Volume 162 Issue 2, 17 May 2004, pp. 81 – 104, Elsevier Science Inc. New York, USA

[7] Bruce G. Buchanan and Edward H. Shortliffe, "Rule Based Expert Systems: The Mycin Experiments of the Stanford Heuristic Programming Project", *Addison-Wesley Longman Publishing Co.*, Inc. Boston, MA, USA, 1984, ISBN: 0201101726.

[8] Cecilia Vallejos de Schatz, Fabio Kurt Schneider, "Intelligent and Expert Systems in Medicine – A Review", *XVIII Congreso Argentino de Bioingeniería SABI 2011 - VII Jornadas de Ingeniería Clínica Mar del PlataSara*, 28 – 30 September, 2011, pp. 326-331

[9] Neuromuscular Disease Division, "What is Neuromuscular Disease?" http://www.neurology.upmc.edu/neuromuscular/patient_info/what.html, Accessed 1 June, 2012

[10] Robert Palisano, Peter Rosenbaum, Stephen Walter, Dianne Russell, Ellen Wood, Barbara Galup, "Development and reliability of a system to classify gross motor function in children with cerebral palsy", *Developmental Medicine & Child Neurology*, Volume 39, Issue 4, pages 214–223, April 1997.

[11] Data and Statistics For Cerebral Palsy, http://www.cdc.gov/ncbddd/cp/data.html

[12] Parkinson Disease Health Center, "What causes Parkinson's Disease?", http://www.webmd.com/parkinsons-disease/parkinsons-causes

[13] All About Parkinson's disease, "What is Parkinson's disease?", http://www.medicalnewstoday.com/info/parkinsons-disease/

[14] U.S. National Library of Medicine , "Muscular Dystrophy, http:// ncbi.nlm.gov/pubmedhealth/PMH002172/

[15] National Multiple Sclerosis Society, "About Multiple Sclerosis", http://www.nationalmssociety.org/about-multiple-sclerosis/index.aspx, Accessed May 1, 2012

[16] John Durkin, "Expert System: Design and Development," New York, *Macmillan Publishing Company*, Inc., 1994, pp. 174-195.

[17] The RETE Algorithm, http://www.cis.temple.edu/~giorgio/cis587/readings/rete.html

[18] Carole Ann, "The RETE algorithm demystified Part 2", http://techondec.wordpress.com/2011/03/14/rete-algorithm-demystified-part-2/